# Observation of three superconducting transitions in the pressurized CDW-bearing compound $TaTe_2$

Jing Guo[1,6], Cheng Huang[1,5], Huixia Luo[2], Huaixin Yang[1], Linlin Wei[1], Shu Cai[1], Yazhou Zhou[1], Hengcan Zhao[1], Xiaodong Li[3], Yanchun Li[3], Ke Yang[4], Aiguo Li[4], Peijie Sun[1], Jianqi Li[1], Qi Wu[1], Robert J Cava[2], and Liling Sun[1,5,6]†

[1]*Institute of Physics, Chinese Academy of Sciences, Beijing 100190, China*

[2]*Department of Chemistry, Princeton University, Princeton, New Jersey 08544, USA*

[3]*Institute of High Energy Physics, Chinese Academy of Science, Beijing 100049, China*

[4]*Shanghai Synchrotron Radiation Facilities, Shanghai Institute of Applied Physics, Chinese Academy of Sciences, Shanghai 201204, China*

[5]*University of Chinese Academy of Sciences, Beijing 100190, China*

[6] *Songshan Lake Materials Laboratory, Dongguan, Guangdong 523808, China*

Transition metal dichalcogenides host a wide variety of lattice and electronic structures, as well as corresponding exotic physical properties, especially under certain tuning conditions. Here, we are the first to report the observation of pressure-induced three superconducting transitions in $TaTe_2$, a charge density wave (CDW) - bearing layered transition-metal dichalcogenide that is metallic but not superconducting at ambient pressure. We find that its CDW state can be easily suppressed upon increasing pressure up to ~ 1 GPa. A superconducting state then emerges from the suppressed CDW state and persists to the pressure about 7 GPa. Unexpectedly, another superconducting state appears at ~ 11 GPa within the same monoclinic (M) structure of its ambient-pressure one. Upon further compression to 21 GPa, a third superconducting state with higher $T_c$ appears from a high-pressure (HP) phase. Our experimental results suggest that the pressure-induced three superconducting transitions in $TaTe_2$ are respectively driven by the suppression of the CDW state, the change of the β angle in the M phase and the transition of M-to-HP phase. These results demonstrate not only the versatile nature of this correlated electron system, but also the first experimental example that shows the pressure-induced evolution from a CDW state to three superconducting states driven by different mechanisms.

In layered transition metal dichalcogenides (TMDs), a charge density wave (CDW) is frequently observed, appearing as a periodic modulation of the electronic charge density [1-6]. Through chemical doping or under applied pressure, many of these compounds show a competition between the CDW ordered state and superconductivity [1,6-11], similar to the behavior that is observed in unconventional superconductors whose superconductivity resides near the boundary of an ordered magnetic state. There are many examples that demonstrate the close connection between superconductivity and ordered state, such as the magnetic orders in iron-based superconductors, the heavy Fermion superconductors and the copper oxide superconductors [12-15]. The same is true for the connection between superconductivity and CDW order.

In this study, we report a new case, finding pressure-induced three superconducting transitions in the non-superconducting CDW-bearing compound $TaTe_2$, revealed by complementary measurements of *in-situ* high pressure electrical resistance, *ac* susceptibility, Hall coefficient, angle dispersive X-ray diffraction (XRD) and Transmission Electron Microscopy (TEM).

The high-quality single crystals were synthesized by the chemical vapor transport method, using iodine as a transport agent [16]. Pressure was generated by two types of high-pressure cells, piston-cylinder and diamond anvil cell, for the low and high pressure measurements. In the low-pressure measurements (up to 2.11 GPa), 7373 oil was used as pressure transmitting medium and pressure was determined by the pressure dependence of $T_c$ of Pb [17] that was placed together with the sample in a Teflon capsule. For the high-pressure measurements, up to ~40 GPa, diamond anvils with 300 μm flats were employed, and NaCl powder was used as the pressure transmitting medium to obtain a quasi-hydrostatic pressure environment. The sample sizes for the low and high pressure measurements were 2mm×1mm×0.2mm and 60μm×60μm×5μm, respectively. High-pressure electrical resistance and Hall coefficient measurements were carried out using a standard four-probe technique and the Van der Paw method [18-20]. High-pressure alternating current (*ac*) susceptibility measurements were conducted using home-made primary/secondary-compensated coils around a diamond anvil [18,21]. High-pressure X-ray diffraction (XRD) measurements were carried out at beamline 4W2 at the Beijing Synchrotron Radiation Facility and at beamline 15U at the Shanghai Synchrotron Radiation Facility. A monochromatic X-ray beam with a wavelength of 0.6199 Å was used and silicon oil was employed as a pressure-transmitting medium. The pressure for all measurements, in a diamond anvil cell, was determined by the ruby fluorescence method [22]. Ambient-pressure TEM observations were performed on a JEOL 2100F transmission electron microscope equipped with a low-temperature sample holder.

First, we performed *in-situ* resistance measurements on the $TaTe_2$ sample in the low pressure range by using a piston-cylinder pressure cell. Figure 1a shows the electrical resistance as a function of temperature down to ~ 0.4 K at different pressures up to 2.11 GPa. It is seen that the resistance of the ambient-pressure sample exhibits an anomaly at ~175 K, due to the CDW transition [23-26]. To better specify this resistance anomaly, we carried out TEM measurements at ambient-pressure at 300 K and 100 K respectively. The electron-diffraction patterns taken along the [-101] direction at two different temperatures are shown in the inset of Fig. 1a. It is seen that a new set of electron-diffraction reflections, which is not present at 300 K, appears in reciprocal space at 100 K, signaling the existence of the CDW order. To know the evolution of the CDW ordered state with pressure, we applied high pressure on the $TaTe_2$ sample and measured the temperature dependence of the resistance in a warming and cooling cycle at different pressures (Fig. 1a and Fig.S1 in Ref. 27-Supplementary Information-SI). A thermal hysteresis in resistivity is observed (Fig.S1 in Ref.27), which is considered to be a typical feature of the CDW transition [23,26]. We found that the formation temperature of the CDW state, $T_{CDW}$, decreases with increasing pressure and then is undetectable at ~ 1.32 GPa (Fig.1a). Moreover, the sample shows a sharp resistance drop at ~ 0.5 K under a pressure of about 0.96 GPa, which is close to the boundary of the suppressed CDW state. A zero-resistance state is achieved at ~ 0.4 K, indicative of a superconducting transition (as shown in the lower right inset of Fig. 1a). The onset temperature of the superconducting transition ($T_c$) increases with further compression up to 2.11 GPa, the maximum pressure of this experimental run. To further characterize the observed superconducting behavior in the pressurized $TaTe_2$, we applied magnetic fields for $TaTe_2$ subjected to 1.75 GPa (Fig. 1b). It is found that its resistance drop shifts to lower temperature with increasing magnetic field and completely vanishes at ~ 900 Oe. These results indicate that the pressure-induced resistance drop is associated with a superconducting transition. We extract the field dependence of midpoint $T_c$ for $TaTe_2$ at 1.75 GPa and estimate the upper critical magnetic field at zero temperature to be about ~ 520 Oe (Fig. 1c).

To investigate the evolution of the observed superconducting state under higher pressure, we carried out high-pressure measurements on the $TaTe_2$ samples in a diamond anvil cell. As shown in Fig. 2a, the superconducting transition with a zero-resistance state is seen at 2.1 GPa and the $T_c$ value slightly shifts to higher temperature at 3.6 GPa. Upon increasing pressure to 5.9 GPa, the sample loses its zero resistance. At 8.3 GPa, the resistance drop is not observable at temperature down to ~ 0.3 K, revealing that the superconducting state is suppressed by high pressure. Interestingly, when we applied pressure up to 11.5 GPa, the resistance drop reemerges and zero resistance is observed at ~ 0.5 K, indicating that a new superconducting state appears (Fig. 2b). The onset transition temperature of this new superconducting state increases with elevating pressure, but the transition becomes broad starting at ~ 15 GPa. Its $T_c$ value shifts to high temperature slightly upon compression to 24 GPa (Fig.2c). At ~ 27.5 GPa, unexpectedly, we found another resistance drop, at ~ 3.5 K. The onset temperature of this resistance drop shifts slightly to higher temperature with increasing pressure (Fig. 2c). We repeated the measurements with new samples in different experiments, and found that the results were reproducible (Fig. 2d-2f and Fig.S2 in Ref.27). To characterize the superconducting transition in the pressure range of 10-22 GPa and 22-40.5GPa, we applied a magnetic field on $TaTe_2$ subjected to 13.6 GPa and 40.5 GPa (Fig. 2g and Fig S3a), and found that the resistance drop shifts to lower temperature with increasing magnetic field and the estimated upper critical magnetic fields at zero temperature are ~ 680 Oe at 13.6 GPa, and ~2850 Oe at 40.5 GPa. The observation of the different upper critical magnetic fields in the second and third superconducting states implies that they may have different natures.

In order to confirm these pressure-induced superconducting transitions, high-pressure *ac* susceptibility measurements were performed in a diamond anvil cell up to 32.6 GPa. The onset temperature of the diamagnetism was observed at 0.5 K, 0.53 K and 0.7 K at pressures of 2.9 GPa, 15.3 GPa and 24.5 GPa, respectively (Fig. 2i). All these results indicate that the pressure-induced resistance drops are attributable to the superconducting transitions of bulk superconductivity. We did not observe a diamagnetic signal from our *ac* susceptibility measurements at the temperature near the onset of the resistance drop in the pressure range that the third superconducting state exists, implying that the superconductivity of the third superconducting state is likely a filament one (Fig.2i). While the diamagnetism observed at 1.15 K and 32.6 GPa should be from the second superconducting state that coexists with the third one at this pressure (see high-pressure XRD data below).

High pressure synchrotron X-ray diffraction (XRD) measurements were performed at the two Synchrotron Radiation Facilities to clarify whether the observed superconducting transitions in pressurized $TaTe_2$ are related to the structure phase transitions (Fig. 3 and Fig.S4 in Ref. 27). The XRD patterns collected at the Beijing Synchrotron Radiation Facility are displayed in Fig. 3a. It is found that all peaks can be indexed well with the monoclinic phase in the *C2/m* space group below 21 GPa, indicating that its ambient-pressure phase of $TaTe_2$ is stable below 21 GPa. However, a new peak appears at ~13 degrees when pressure is increased to 21.3 GPa (Fig.3a) and 23 GPa(Fig.S4 in Ref. 27), and its intensity increases with pressure, indicating that the pressure induces a new structural phase transition. Since the new phase cannot be exactly refined by a single peak, here we have to define it as the high pressure (HP) phase. We found that this HP phase coexists with the ambient-pressure M phase under pressure up to 40 GPa (Fig.S3 in Ref.27). Based on our XRD results, we extract the pressure dependence of lattice parameters and volume (Fig. 3b-3d). It is seen that the pressure-dependent lattice constants and volumes determined from the two independent diffraction measurements at the two different synchrotron x-ray sources are in good agreement with each other, and the lattice constants and volume do not exhibit any discontinuities with increasing pressure to 40 GPa. These results demonstrate that the ambient-pressure M phase is stable under pressure to 40 GPa, which indicates that the HP phase is developed from the matrix of its ambient-pressure M phase.

We summarize the high-pressure transport results and structure information for $TaTe_2$ in Fig. 4a. It is found that the first superconducting (SC-I) state emerges from a suppressed CDW state, vanishes at ~7 GPa, and that the second superconducting (SC-II) state appears at ~11 GPa, followed by the appearance of the third superconducting (SC-III) state at ~21 GPa. The crystal structure of the SC-I state and the SC-II state reside in the ambient-pressure M phase, while the SC-III state is likely to be within the HP phase, because the SC-III phase appears just when the HP phase forms at ~21 GPa and the $T_c$ value of the $TaTe_2$ sample shows a jump at ~21 GPa, which is often seen in many of the phase-transition-induced superconducting state [28]. Therefore, we propose that the formation of the HP-phase is responsible for the emergence of SC-III phase. Moreover, our proposal about that superconductivity of the SC-III state may be a filament-like one is also supported by our XRD results, in which only one XRD peak of the HP phase appears on the matrix of the M phase within the pressure range it exists. Certainly, the nature of the SC-III state calls for further investigations both from experimental and theoretical sides.

To understand the evolution of the pressure-induced multiple superconducting transitions from the CDW ordered state, we performed high-pressure Hall resistance and magnetoresistance measurements on the $TaTe_2$ sample by sweeping the magnetic field perpendicular to the *ab* plane at two temperatures (4K and 10 K) and various pressures. The pressure dependence of the Hall coefficient ($R_H$) and the $MR$% ($MR$ is defined as [($R(7T)$-$R(0T)$/$R(0T)$]×100%) are illustrated in Fig. 4b and the SI. At ambient pressure, the Hall coefficient ($R_H$) displays a positive sign both at 4 K and 10 K, implying that hole-carriers are dominant. Meanwhile, the sample displays a positive magnetoresistance effect ($MR$%= 62). Within the pressure range of the CDW ordered state, $R_H$ and $MR$% dramatically decrease with increasing pressure, suggesting that the role of electron carriers is enhanced by applying pressure. At ~1 GPa, the SC-I state emerges from the suppressed CDW order state, in accordance with the common picture seen in layered transition metal dichalcogenides [6,8,9,29], implying that the superconducting electrons in the SC-I state are released from the CDW state. The evolution of $T_c$ with pressure in SC-I state forms a dome-like shape with a maximum $T_c$ of 0.7 K at ~3.5 GPa. While, in the pressure range of ~11-21 GPa, the SC-II state presents without a structural phase transition. We found that the values of the Hall coefficient $R_H$ at 4 K and 10 K remain nearly constant in the pressure range where the SC-II state exists, suggesting that the SC-II state develops from a semimetal state [30]. To further investigate the origin the SC-II state, we extracted the pressure dependence of the angle β that represents the lattice deformation degree of the monoclinic phase and found that the value of β angle is ~ 111 degree at ambient pressure, and it is reduced with increasing pressure, reaches to 108.5 degree at ~12.5 GPa where the SC-II state presents. We suggest that reducing β angle by ~2.5 degrees and more can induce the appearance of the SC-II state. At ~ 21 GPa, the β angle starts to show an increase with pressure, which is likely associated with the presence of the HP phase developed from the matrix of the monoclinic phase.

We propose that the semimetal state may be associated with the peculiar positive magnetoresistance state that is observed (Fig.4b). Such behavior is reminiscent of what has been seen in the pressurized Weyl semimetal $WTe_2$, which displays a large positive magnetoresistance at ambient pressure [31]. High pressure studies on $WTe_2$ found that as long as the positive magnetoresistance effect prevails, no superconductivity is present, *i.e.* the superconductivity only appears when the positive magnetoresistance effect is completely suppressed [19, 32]. For pressurized $TaTe_2$, we found that the SC-III state with higher superconducting transition temperature emerges at pressures above ~21 GPa, a pressure where a new structure sets in (Fig.3), meanwhile, the positive magnetoresistance suddenly disappears and the Hall coefficient shows a drop starting at ~21 GPa (Fig.4b). Based on these experimental results, we propose that the HP phase emerging at ~ 21 GPa manifests a topology change of the Fermi surface, which in turn tips the electronic structure at the Fermi level in favor of superconductivity.

In summary, we report the observation of pressure-induced three superconducting transitions in the CDW-bearing compound $TaTe_2$. We propose the possible reasons for these three superconducting states as follows: the suppression of the CDW ordered state and the enhancement of superconducting carrier density lead to the appearance of the SC-I state; the critical β angle of the M phase drives the emergence of the SC-II state; and the pressure-induced structural phase transition gives rise to the development of the SC-III state. To the best of our knowledge, this is the only experimental example that three different superconducting transitions can be tuned out by pressure from one compound with a homogenies lattice structure. These results are expected to shed new insight on the underlying superconducting mechanisms in the correlated electron systems of the transition metal ditellurides, and even in the other unconventional superconducting compounds.

**Acknowledgements**

This work in China was supported by the NSF of China (Grant Numbers Grants No. U2032214, 12122414, 12104487 and 12004419), the National Key Research and Development Program of China (Grant No. 2017YFA0302900 and 2017YFA0303103), and the Strategic Priority Research Program (B) of the Chinese Academy of Sciences (Grant No. XDB25000000). We thank the support from the Users with Excellence Program of Hefei Science Center CAS (2020HSC-UE015). Part of the work is supported by the Synergic Extreme Condition User System. J. G. is grateful for support from the Youth Innovation Promotion Association of the CAS (2019008). The work at Princeton was supported by the US Department of Energy, Division of Basic Energy Sciences, grant DE-FG02-98ER45706.

† Correspondence and requests for materials should be addressed to L.S. (llsun@iphy.ac.cn)

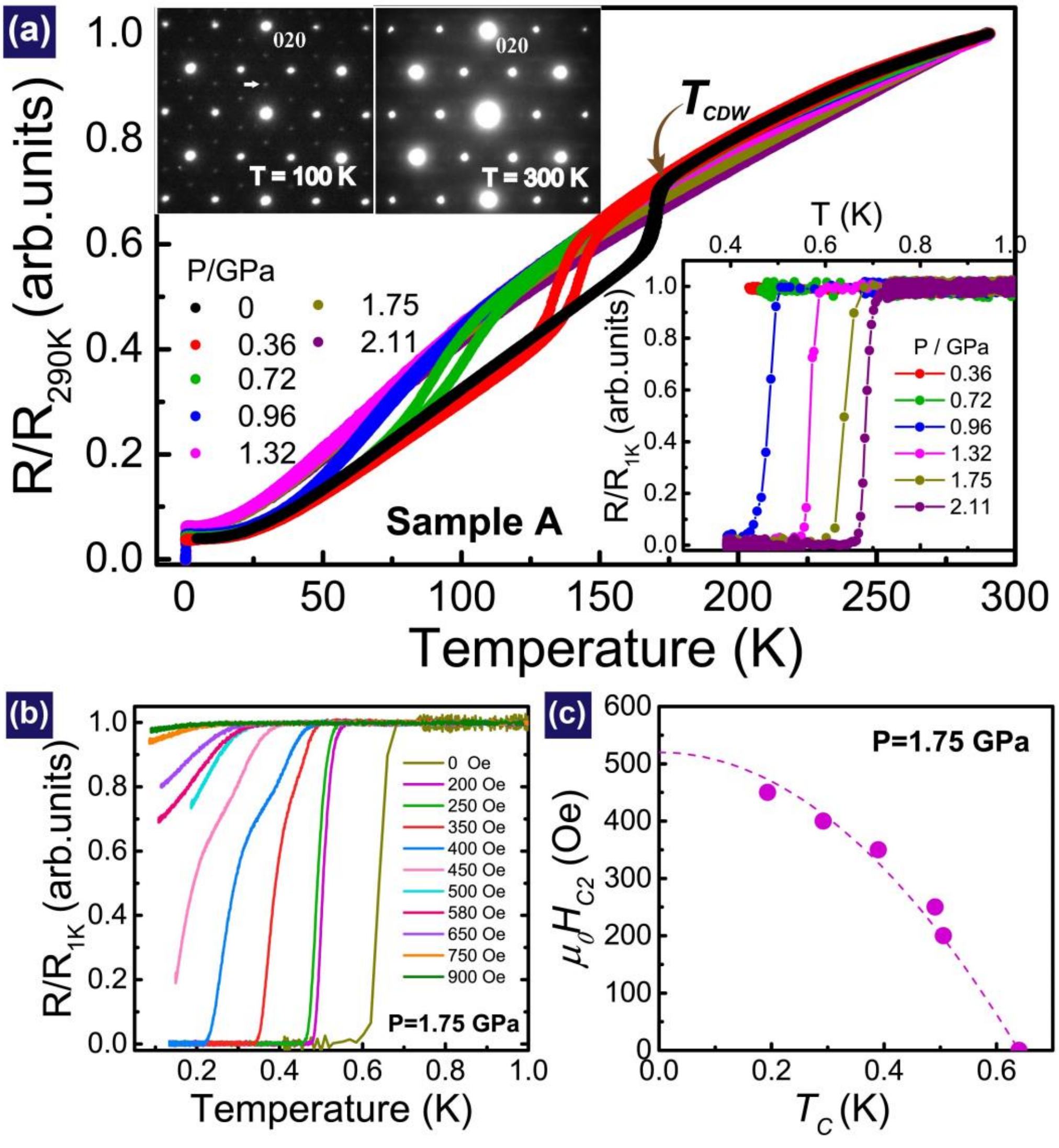

**Fig. 1 Characterizations for $TaTe_2$** single crystal (a) The normalized resistance as a function of temperature at pressures below 2.11GPa. The upper left inset display

electron-diffraction patterns taken along the [-101] zone axis direction at 100 K and 300 K, respectively; the superstructure due to the existence of the CDW state at low temperature is indicated by an arrow. The lower right inset displays enlarged view of the normalized resistance versus temperature below 1 K. (b) The temperature dependence of the normalized resistance under different magnetic fields at 1.75 GPa. (c) $H_{C2}$ as a function of temperature. The dashed line represents the Ginzburg-Landau (GL) fits.

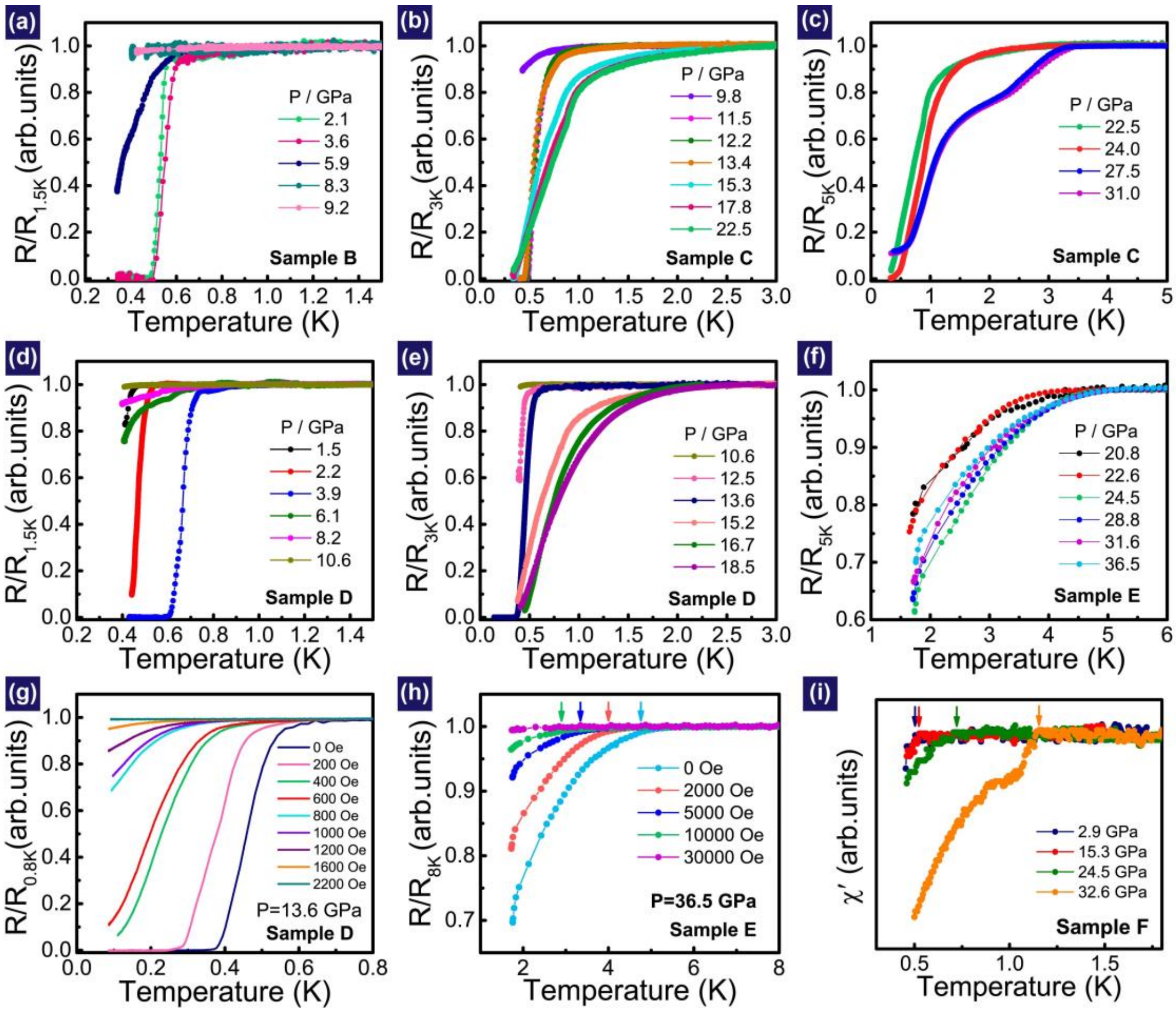

**Fig. 2 Electronic resistance and diamagnetism measurements for pressurized $TaTe_2$** (a) The normalized resistance as a function of temperature for the sample B in the pressure range of 2.1-9.2 GPa. (b)-(c) for the sample C in the pressure range of 9.8-22.5 GPa and 22.5-31.0 GPa, respectively. (d)-(e) for the sample D obtained in the pressure range of 1.5–10.6 GPa and 10.6-18.5 GPa, respectively. (f) for the sample E in the pressures range of 20.8 to 36.5 GPa. (g)-(h) The temperature dependence of the normalized resistance under different magnetic fields at 13.6 GPa for the sample D and at 36.5 GPa for the sample E. (i) The real part of the alternating-current (*ac*) susceptibility ($\chi'$) as a function of temperature for the sample F at different pressures. The arrows denote the temperatures of the superconducting transition.

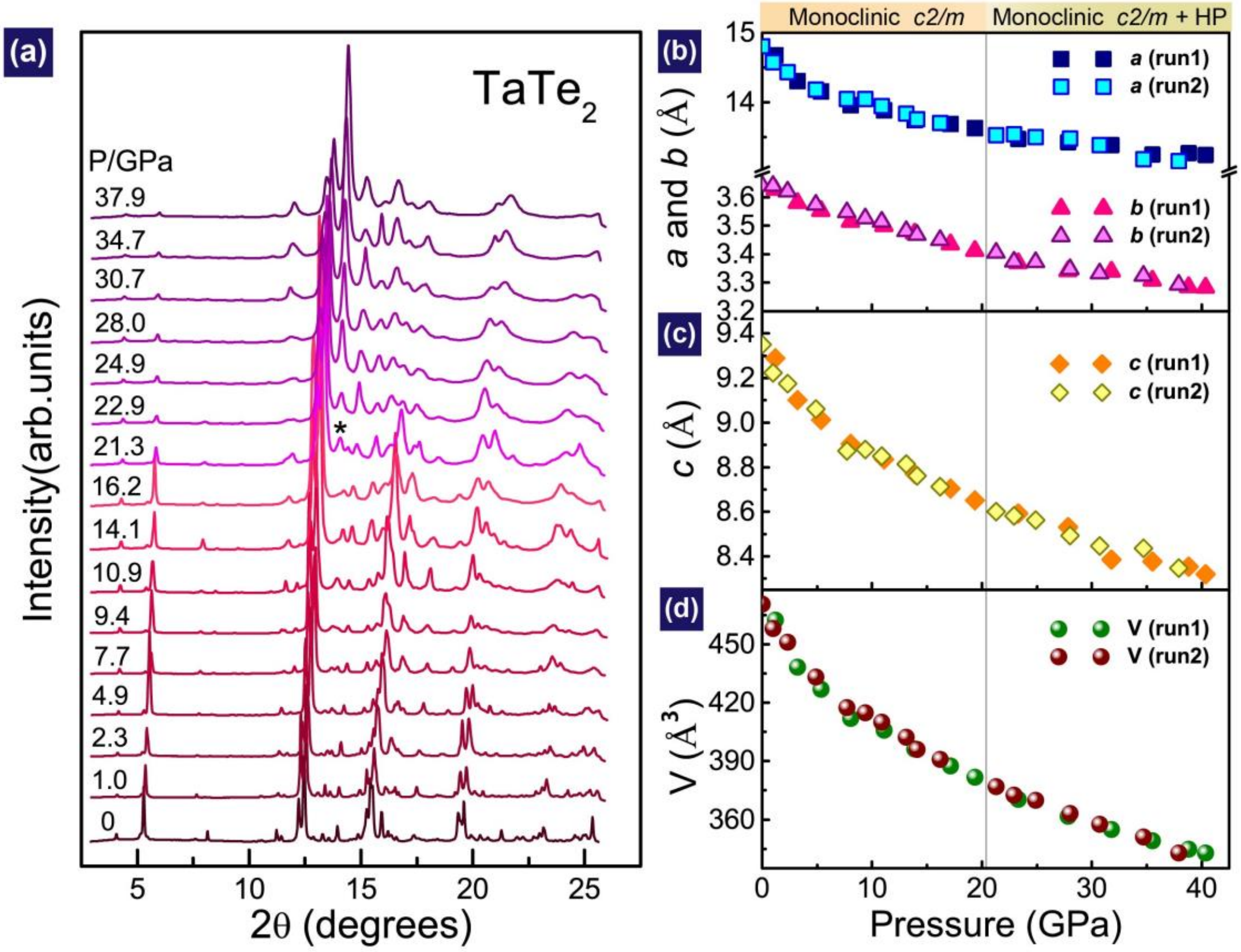

**Fig. 3 Structural characterization of $TaTe_2$** (a) The X-ray diffraction patterns obtained at different pressures. A new peak beginning at ~21 GPa is observed, indicated by a star, suggesting that at this pressure a new structure phase emerges on the matrix of the monoclinic phase. (b)-(d) The pressure dependence of the lattice constants (*a, b* and *c*) and unit cell volume (*V*) obtained from two independent measurements at two different synchrotron sources.

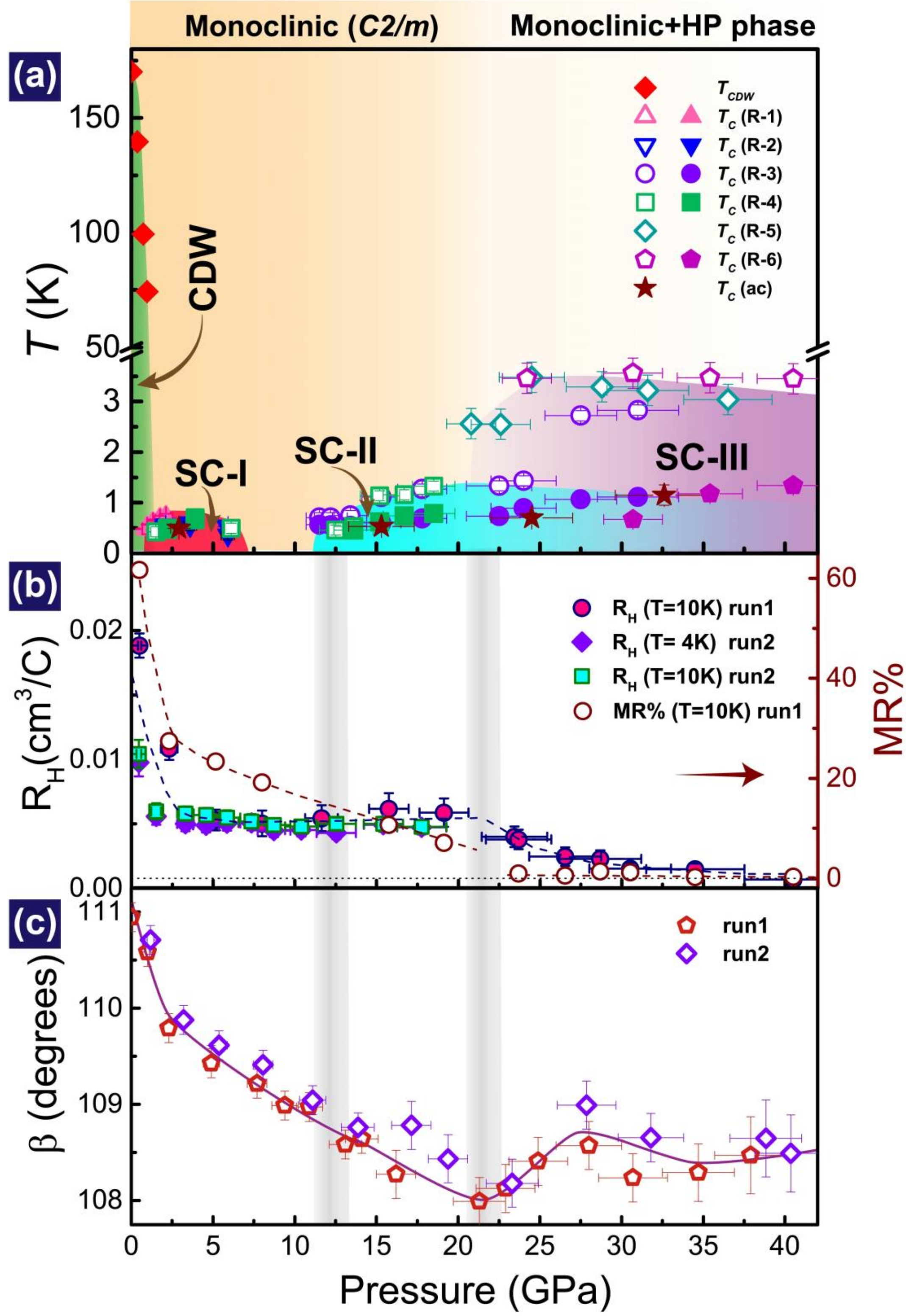

**Fig. 4 Summary of the experimental observations on $TaTe_2$** (a) Pressure-Temperature phase diagram combined with structure information for $TaTe_2$.

$T_{CDW}$ represents the formation temperature of the CDW state. $T_c$(R) represents the superconducting transition temperature determined by the resistance measurements. The open and solid symbols represent the superconducting transition temperatures that were defined as temperatures where the resistance falls to 90% and 50% of the normal state value, respectively. $T_c$(ac) represents the superconducting transition temperature determined by *ac* susceptibility measurements. (b) The pressure dependent Hall coefficient ($R_H$) measured at 4 K and 10 K (left axis). The magnetoresistance ($MR$) as a function of pressure measured at 10 K (right axis), here $MR\%=[R(7T)-R(0)]/R(0T)\times100\%$. (c) The β angle of the monoclinic structure as a function of pressure.